\documentclass{article}

\title{Container Profiler: Profiling Resource Utilization of Containerized Big Data Pipelines}

\author{
    Varik Hoang $^*$, Ling-Hong Hung $^*$, David Perez, Huazeng Deng,\\
    Raymond Schooley, Niharika Arumilli, Ka Yee Yeung, Wes J. Lloyd $^1$ \\
    School of Engineering and Technology, University of Washington\\
    Box 358426, Tacoma, WA 98402, USA
    }

\footnotetext{Contributed equally}
\footnotetext{Corresponding author: wlloyd@uw.edu}

\date{\today}

\usepackage{hyperref}
\usepackage{mathtools}
\usepackage{graphicx} 
\usepackage{bmpsize}
\usepackage{adjustbox}
\usepackage{float}
\usepackage{caption}
\usepackage[
backend=biber,
style=numeric,
]{biblatex}

\begin{document}

\maketitle


\begin{abstract}
\noindent
\textbf{Background} This paper presents the {\em Container Profiler}, a software tool that measures and records the resource usage of any containerized task.  Our tool profiles the CPU, memory, disk, and network utilization of containerized tasks collecting over fifty Linux operating system metrics at the virtual machine, container, and process levels.  The {\em Container Profiler} supports performing time series profiling at a configurable sampling interval to enable continuous monitoring of the resources consumed by containerized tasks and pipelines.

\noindent
\textbf{Results}  To investigate the utility of the {\em Container Profiler}, we profile the resource utilization requirements of a multi-stage bioinformatics analytical pipeline (RNA sequencing using unique molecular identifiers). We examine profiling metrics to assess patterns of CPU, disk, and network resource utilization across the different stages of the pipeline. We also quantify the profiling overhead of our Container Profiler tool to assess the impact of profiling a running pipeline with different levels of profiling granularity verifying that impacts are negligible.

\noindent
\textbf{Conclusions} The {\em Container Profiler} provides a useful tool that can be used to continuously monitor the resource consumption of long and complex containerized applications that run locally or on the cloud. This can help identify bottlenecks where more resources are needed to improve performance.

\end{abstract}


\newpage

\section{Introduction}

\subsection{Background}

Large-scale and diverse biomedical data have been generated to advance the understanding of biological mechanisms. Interpreting these data typically includes multiple analytical steps, each of which consists of different computational methods and software tools. An analytical {\em pipeline} (or {\em workflow}) is a sequence of computational tasks used to process and analyze specific biomedical data.
Each analytical step in a pipeline can potentially require a different set of applications, libraries, and software dependencies. As a result, software containers that encapsulate executables with their dependencies have become popular to facilitate the deployment of complicated pipelines and to enhance their reproducibility ~\cite{o2017dockstore, da2017biocontainers}. 
Additionally, different analytical steps in a pipeline could have different computing resource requirements.
In particular, many bioinformatics pipelines consist of one or more computationally intensive steps stemming from their operation on large datasets requiring significant CPU, memory, network, and disk resources. As an example, the alignment step in a RNA sequencing pipeline typically requires relatively more CPU and memory resources than other steps, while the data download step typically requires more disk I/O and network resources.

Cloud computing has emerged as a solution that can provide the necessary resources needed for computationally intensive bioinformatics analyses ~\cite{dai2012bioinformatics, schadt2010computational, schadt2011cloud, lau2017cancer, reynolds2017isb, afgan2011harnessing, birger2017firecloud}. However, deployment of analytical pipelines using Infrastructure-as-a-Service (IaaS) cloud platforms requires selecting the appropriate type and quantity of virtual machines (VMs) to address performance goals while balancing hosting costs. Cloud resource type selection is presently complicated by the rapidly growing number of available VM instance types and pricing models offered by public cloud providers.  For example, the Amazon, Microsoft, and Google public clouds presently offer hundreds of different VM types under different pricing models. Further, Google allows users to create custom VM types with unique combinations of CPUs, memory, and disk capacity.  These cloud VMs are available directly, or through various container platforms. Determining the best cloud deployment requires understanding the resource requirements of the pipeline. 

\subsection{Our Contributions}
This paper presents the {\em Container Profiler}, a tool that supports profiling the computational resources utilized by software within a Docker container. Our tool is simple, easy-to-use, and can record the resource utilization for any Dockerized computational job.   Understanding fine-grained resource utilization of containerized computational tasks can help identify resource bottlenecks and inform the choice of optimal cloud deployment.  The {\em Container Profiler} collects metrics to characterize the CPU, memory, disk, and network resource utilization at the VM, container, and process level. In addition, the {\em Container Profiler} supports time-series graphing enabling the visualization and monitoring of resource utilization of containerized tasks and pipelines.  We present a case study using a multi-stage containerized bioinformatics pipeline that analyzes the unique molecular identifiers (UMI) of RNA sequencing data to illustrate the utility of our tools.
\\

\textbf{Key Points}
\begin{itemize}
\item We present the {\em Container Profiler} a tool that enables profiling the resource utilization of any script or container-based task on Linux.
\item The {\em Container Profiler} collects CPU, memory, disk, and network resource utilization metrics at the virtual machine, container, and process levels.
\item The {\em Container Profiler} supports delta and time series resource utilization profiling at an adjustable time interval (e.g. 1-second) supporting monitoring and graphing of resource utilization enabling time series analysis to help identify performance bottlenecks for any Linux-based computational task.
\item The {\em Container Profiler} can profile complex containerized computational jobs such as bioinformatics pipelines where multiple individual containers are used to implement specific steps. 
\item The {\em Container Profiler} is provided as a container which can merge with any existing container or used separately to profile independent Linux scripts or executables to characterize task resource utilization locally or on the cloud. 
\item We illustrate how different resources are required when performing different steps of a containerized pipeline analyzing unique molecular identifiers (UMI) RNA sequencing data.
\end{itemize}

\section{Related Work}

Cloud computing has been used to process massive RNA sequencing (RNA-seq) datasets~\cite{tatlow2016cloud, lachmann2018massive}.  These pipelines typically consist of multiple computational tasks, where not all tasks necessarily have the same resource requirements. Tatlow {\em et al.} studied the performance and cost profiles for processing large-scale RNA-seq data using pre-emptible virtual machines (VMs) on the Google Cloud Platform~\cite{tatlow2016cloud}. The authors collected resource utilization metrics to characterize user and system vCPU utilization, memory usage, disk activity, and network activity for the different computational stages of the RNA-seq pipeline. Tatlow {\em et al.} observed how resource utilization can vary dramatically across different processing tasks in the pipeline, while demonstrating that resource profiling can help to identify resource requirements of unique pipeline stages. Juve {\em et al.} developed a pair of tools called wfprof (pipeline profiling) to collect and summarize performance metrics for diverse scientific pipelines from multiple domains including bioinformatics~\cite{juve2013characterizing}.  Wfprof consists of two tools, ioprof to measure process I/O, and pprof that characterizes process runtime, memory usage, and CPU utilization.  These tools accomplish profiling at the machine level primarily by analyzing process level resource utilization, and they do not focus on profiling containerized pipelines, nor do they collect container specific metrics.

Tyryshkina, Coraor, and Nekrutenko leveraged coarse grained resource utilization data from historical job runs collected over 5 years on the Galaxy platform to estimate the required CPU time and memory to improve task scheduling~\cite{tyryshkina2019predicting}.  Galaxy, a scientific workflow, data integration, data analysis, persistence, and publishing platform was initially developed for genomics research and is now considered largely domain agnostic and is used for processing general bioinformatics pipelines. The authors identified the challenge of determining the appropriate amount of memory and processing resources for scheduling bioinformatics analyses at scale. The majority of metrics in the study consisted of metadata regarding job configurations. Assessing the utility of using fine grained operating system metrics as with the Container Profiler to profile resource utilization of genomics pipelines was not the focus. This effort considered many older jobs that ran on Galaxy where containers were not used thus they lacked container based metrics.

Outside bioinformatics, Weingartner et al. highlight the importance of profiling resource requirements of applications for deployment in the cloud to improve resource allocation and forecast performance~\cite{weingartner2015cloud}. Brendan Gregg described the USE method (Utilization, Saturation, and Errors) as a tool to diagnose performance bottlenecks~\cite{gregg2013thinking}. Gregg's method involves checking utilization of every resource involved in the system including CPUs, disks, memory, and more to identify saturation and errors.  Lloyd {\em et al.} provided a virtual machine manager known as VM-scaler that integrated resource utilization profiling of software deployments to Infrastructure-as-a-Service (IaaS) cloud VMs~\cite{lloyd2014virtual}.  VM-scaler focused on the management and profiling of cloud infrastructure used to host environmental modeling web services.  This work was later extended by building resource utilization models to enable identifying the most cost effective cloud VM types to host environmental modeling web service workloads without sacrificing runtime or throughput~\cite{lloyd2015demystifying}.  This effort demonstrated a cost variance of ~25\% for hosting these workloads across different VM types on the Amazon Elastic Compute Cloud (EC2) while identifying potential for cost savings up to \$25,000 for 10,000 hours of compute time.

To characterize resource requirements of containerized tasks and pipelines, a variety of commercial and open source tools exist. The vast majority of the available tools, however, require the setup and maintenance of a complete monitoring application including a time-series database and web application server.  ~\cite{cp13} These monitoring applications require dedicated infrastructure (i.e. servers and/or virtual machines) to run always-on daemons. Many of these tools are also oriented towards monitoring entire container clusters (e.g. Kubernetes). Access to such cluster-level monitoring tools is often restricted organizationally to system administrators and privileged users and not made freely available to any user.  For container profiling, there are far fewer solutions that enable a user to easily profile the resource utilization of containerized tasks or pipelines on a local computer or personal cloud VM with minimal effort and expertise. The lack of lightweight easy-to-use developer tools that require no setup or maintenance of a permanent monitoring application and/or database server is what motivated the creation of the {\em Container Profiler}. A related tool, CMonitor has been developed to support similar goals of lightweight container profiling without setup of a full monitoring application ~\cite{cp14, ji2019cmonitor}. CMonitor is installed and run on the host and is used to profile host metrics in addition to container metrics as the tool is not focused specifically on profiling a containerized task or pipeline.  CMonitor, however, runs as an external tool which requires the user to posses detailed information about the host's operating system, runtime configuration, and Docker setup.  Additionally CMonitor does not support container profiling of ARM-based Linux VMs or servers.  These systems are of interest with the advent of low-cost compute-optimized VMs based on the Graviton series of ARM CPUs (e.g. c6g and c7g) on Amazon EC2 ~\cite{arm1,arm2,arm3}.  These VMs offer performance improvements and cost savings of interest for executing bioinformatics pipelines.  
CMonitor is installed as a package requiring several dependencies. 

\section{Container Profiler: Overview}

The {\em Container Profiler} tool supports profiling resource utilization including CPU, memory, disk, and network metrics of containerized tasks.  Resource utilization metrics are obtained across three levels: virtual machine (VM)/host, container, and process.  Our implementation leverages facilities provided by the Linux operating system that is integral with Docker containers. Development and testing of the {\em Container Profiler} described in this paper was completed using Debian-based Ubuntu Linux.

\begin{figure*}[ht]
\centering
\caption{Overview summarizing resource utilization metrics (61 total) collected by the {\em Container Profiler} across three levels (i.e. host/VM, container, and process level) and four categories (i.e. CPU, memory, network, and disk).  Process level metrics are depicted by red and prefaced with lower case "p", container level metrics by yellow and prefaced with lower case "c", and host/VM level metrics by blue and prefaced with lower case "v".}
\includegraphics[width=\textwidth]{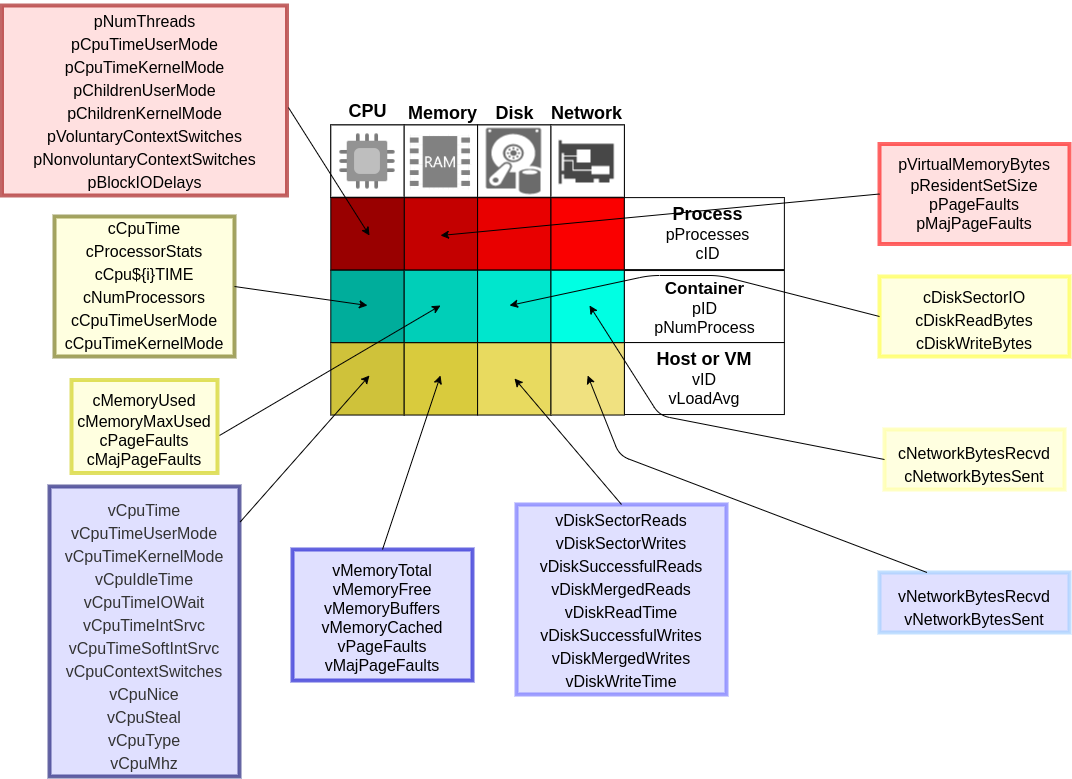}
\label{fig:association}
\end{figure*}

The {\em Container Profiler} collects information from the Linux \texttt{/proc} and \\
\texttt{/sys/fs/cgroup} file systems while a workload is running inside a container on the host machine.  To support collecting metrics the {\em Container Profiler} is implemented using Python3 while leveraging psutil, a cross-platform library for retrieving information on running processes and system utilization ~\cite{psutil}.  The host machine could be a physical computer such as a laptop or a virtual machine (VM) in the public cloud.  The workload being profiled can be any job capable of running inside a Docker container.  Figure~\ref{fig:association} provides an overview of the various metrics collected by the {\em Container Profiler}.\\

\begin{table*}[htb]
  \caption{Selected CPU, disk, and network utilization metrics profiled at the VM/host level.}
  \label{tab:vmmetric}
  \begin{adjustbox}{width=\textwidth}
  \begin{tabular}{l|c|c}
    \hline 
    \textbf{Metric} & \textbf{Description} & \textbf{Source} \\
    \hline
    \hline
    vCpuTimeUserMode & Time the CPU spent executing in user mode  & /proc/stat \\
    \hline
    vCpuTimeKernelMode & Time the CPU spent executing in kernel mode & /proc/stat\\
    \hline
    vCpuIdleTime & Time the CPU was idle & /proc/stat \\
    \hline
    vCpuTimeIOWait & Time the CPU waits for I/O to complete  & /proc/stat \\
    \hline
    vCpuContextSwitches & The total number of context switches across all CPU cores & /proc/stat \\
    \hline
    vDiskSectorReads & Number of sector reads & /proc/diskstats \\
    \hline
    vDiskSectorWrites & Number of sectors writes & /proc/diskstats \\
    \hline
    vDiskReadTime & Time spent reading & /proc/diskstats \\
    \hline
    vDiskWriteTime & Time spent writing & /proc/diskstats \\
    \hline
    vNetworkBytesRecvd & Network Bytes received & /proc/net/dev \\
    \hline
    vNetworkBytesSent & Network Bytes written & /proc/net/dev \\
    \hline
  \end{tabular}
  \end{adjustbox}
\end{table*}

\begin{table}[htb]
  \caption{Selected CPU, disk, and network utilization metrics profiled at the container level.}
  \label{tab:containermetric}
  \begin{adjustbox}{width=\textwidth}
  \begin{tabular}{l|c|c}
    \hline 
    \textbf{Metric} & \textbf{Description} & \textbf{Source} \\
    \hline
    \hline
    cCpuTimeUserMode & CPU time consumed by tasks in user mode  & /sys/fs/cgroup/cpuacct/cpuacct.stat \\
    \hline
    cCpuTimeKernelMode & CPU time consumed by tasks in kernel mode & /sys/fs/cgroup/cpuacct/cpuacct.stat\\
    \hline
    cDiskSectorIO & Number of sectors transferred to or from specific devices & /sys/fs/cgroup/blkio/blkio.sectors \\
    \hline
    cDiskReadBytes & Number of bytes transferred from specific devices  & /sys/fs/cgroup/blkio/blkio.throttle.io\_service\_bytes \\
    \hline
    cDiskWriteBytes & Number of bytes transferred to specific devices & /sys/fs/cgroup/blkio/blkio.throttle.io\_service\_bytes \\
    \hline
    cNetworkBytesRecvd & The number of bytes each interface has received & /proc/net/dev \\
    \hline
    cNetworkBytesSent & The number of bytes each interface has sent & /proc/net/dev \\
     \hline
  \end{tabular}
  \end{adjustbox}
\end{table}

\begin{table*}[h!t]
  \renewcommand{\arraystretch}{1.3}
  \centering
  \caption{List of important metrics for profiling process resource utilization.}
  \label{tab:processmetric}
  \begin{adjustbox}{width=\textwidth}
  \begin{tabular}{l|c|c}
    \hline 
    \textbf{Metric} & \textbf{Description} & \textbf{Source} \\
    \hline
    \hline
    pCpuTimeUserMode & Amount of time that this process has been scheduled in user mode  & /proc/[pid]/stat \\
    \hline
    pCpuTimeKernelMode & Amount of time that this process has been scheduled in kernel mode & /proc/[pid]/stat\\
    \hline
    pVoluntaryContextSwitches & Number of voluntary context switches & /proc/[pid]/status \\
    \hline
    pNonvoluntaryContextSwitches & Number of involuntary context switches  & /proc/[pid]/status \\
    \hline
    pBlockIODelays & Aggregated block I/O delays & /proc/[pid]/stat \\
    \hline
    pResidentSetSize & Number of pages the process has in real memory & /proc/[pid]/stat \\
    \hline
  \end{tabular}
  \end{adjustbox}  
\end{table*}

\textbf{Host-Level Metrics:}  Host/VM level resource utilization metrics are obtained from the Linux \texttt{/proc} virtual filesystem using psutil.  The \texttt{/proc} filesystem is a virtual filesystem that consists of dynamically generated files produced on demand by the Linux operating system kernel providing an immense amount of data regarding the state of the system~\cite{cp12}.  Files in the \texttt{/proc} filesystem are generated at access time from metadata maintained by Linux to describe current resource utilization, devices, and hardware configuration as managed by the Linux kernel. The {\em Container Profiler} queries the \texttt{/proc} filesystem directly and by using the psutil library at regular time intervals to obtain resource utilization metrics. Documentation regarding the Linux \texttt{/proc} filesystem is found on the \texttt{/proc} Linux manual pages~\cite{cp12} though other references provide more detailed descriptions of available metadata: ~\cite{cp1,cp2,cp3,cp4,cp5,cp6,cp7,cp8,cp9,cp10,cp11}. User-mode and kernel-mode CPU utilization metrics can be obtained found in the \texttt{/proc/stat} file. Table~\ref{tab:vmmetric} provides a subset of CPU, disk, and network utilization metrics profiled by the {\em Container Profiler} at the VM/host level.

\textbf{Container-Level Metrics:}  Docker relies on the Linux \texttt{cgroup} and \texttt{namespace} features to facilitate the aggregation of a set of Linux processes together to form a container.  \texttt{Cgroups} were originally added to the Linux operating system to provide system administrators with the ability to dynamically control hardware resources for a set of related Linux processes~\cite{cgroups}.  Linux control groups (\texttt{cgroups}) provide a kernel feature to both limit and monitor total resource utilization of containers. Docker leverages \texttt{cgroups} for resource management to restrict hardware access to the underlying host machine to facilitate sharing when multiple containers share the host. Linux subsystems such as CPU and memory are attached to a \texttt{cgroup} enabling the ability to control resources of the \texttt{cgroup}. Resource utilization of \texttt{cgroup} processes is aggregated for reporting purposes under the \texttt{/sys/fs/cgroup} virtual filesystem and we leverage this filesystem to obtain container-level metrics in the {\em Container Profiler}. \texttt{Cgroup} files provide aggregated resource utilization statistics describing all of the processes inside a container.  Container-level metrics are not available from psutil. 
As a profiling example, a container's CPU utilization statistics can be obtained from \texttt{/sys/fs/cgroups/cpuacct/cpuacct.stat}. Table~\ref{tab:containermetric} describes a subset of the CPU, disk, and network utilization metrics profiled at the container level by the {\em Container Profiler}.

\textbf{Process-Level Metrics:}  The {\em Container Profiler} also supports profiling the resource utilization for each process running inside a container. The {\em Container Profiler} leverages support from the psutil library to capture process level metrics from Linux. Table~\ref{tab:processmetric} describes a subset of the process-level metrics collected by the {\em Container Profiler} to profile resource utilization of container processes.

Resource utilization data collected at the VM/host, container, and process level allows characterization of resource use with increasingly greater isolation.  Host-level resource metrics for example, do not isolate background processes.  This could lead to variance in measurements as background processes on the host machine outside the container may be randomly present. Profiling at the container level allows fine-grained resource profiling of ONLY the resources used by the containerized task or pipeline.  Finally, profiling at the process level allows very fine-grained profiling so that resource bottlenecks can be attributed to the specific activities or tasks. The ability of the {\em Container Profiler} to characterize resource utilization at multiple levels enables high observability of the resource requirements of computational tasks.  This observability can be crucial to improving job deployments to cloud platforms to alleviate performance bottlenecks and optimize performance and cost.

\section{Results}
We demonstrate the Container Profiler using unique molecular identifier (UMI) RNA sequencing data generated by the LINCS Drug Toxicity Signature (DToxS) Generation Center at Icahn School of Medicine at Mount Sinai in New York~\cite{umi-xiong}. The scripts and supporting files for the analytical pipeline to analyse this originated from the Broad Institute~\cite{Soumillon003236}. In addition to downloading the datasets, there are 3 other stages. The first stage is a demultiplexing or split step that sorts the reads using a sequence barcode to identify the originating sample. The second stage aligns the reads to a human reference sequence to identify the gene that produced the transcript. The final stage is the "merge" step which counts all the aligned reads to identify the number of transcripts produced by each gene. The unique molecular identifier (UMI) sequence is used to filter out reads that arise from duplication during the sample preparation process. In the original pipeline, only the most CPU intensive part of the pipeline, the alignment step, was optimized and executed in parallel. We further optimized the split and align steps in the original pipeline~\cite{Soumillon003236} to decrease the running time from 29 to 3.5 hours in our previous work~\cite{hung2019holistic}. We also encapsulated each step in the pipeline in separate Docker containers to facilitate deployment and ensure reproducibility.

\begin{figure}[H]
\caption{CPU utilization graph for the four stages (e.g. download, split, align, and merge) of the UMI RNA-seq pipeline.  This graph depicts the percentage of CPU utilization in each CPU mode.  CpuUsr (shown in green) captures time the pipeline spent executing its source code.  CpuKrn (shown in yellow) captures time when the processor executed code in the Linux kernel. Typically the kernel is invoked to support disk and network I/O which are considered privileged operations.  CpuIdle (shown in blue) is unused time across the 8 available CPU cores throughout each stage.  CpuIdle time is common when waiting for disk or network I/O to complete. High CpuIdle time during computational stages indicates potential for performance optimization with better parallelization of code.  CpuIOWait (shown in maroon) depicts CPU time where the pipeline was waiting for I/O (disk or network) to complete.  cpuSftIntSrvc (shown in magneta) is time spent handling soft interrupts.  Soft interrupts commonly occur with network I/O.} 
\label{fig:rugraphg}
\centering
\includegraphics[width=\textwidth]{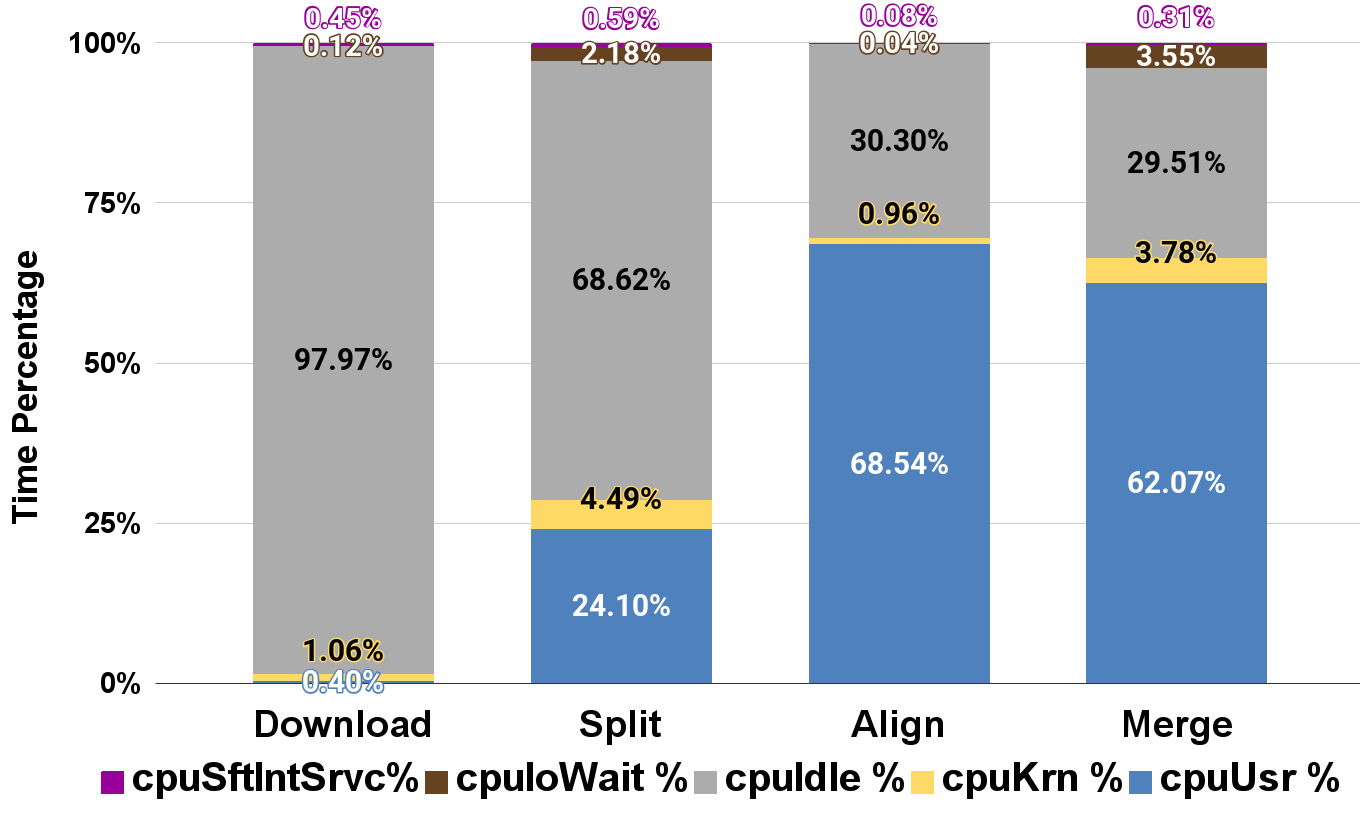}
\end{figure}

We leveraged the UMI RNA-sequencing pipeline as our case study for the {\em Container Profiler} as each stage of the RNA-seq pipeline exhibits different resource utilization characteristics. Specifically, the dataset download stage is limited by the network capacity. The split stage writes many files and is limited by the speed of disk writes. The alignment stage is performed by multiple CPU-intensive processes and performance is primarily limited by the CPU. However, it is possible that available memory capacity will limit the performance in some circumstances. The final merge stage involves reading many files in parallel, consuming both memory and CPU resources depending on the number of threads used.

\begin{figure}[H]
\caption{Output graphs comparing Container and VM (host) level metrics over time for a multi-stage RNA sequencing data pipeline. Four output graphs are shown: disk writes (top left), CPU usage (top right), network usage (bottom left) and memory usage (bottom right). In each graph, the container level metrics are shown in red and the VM (host) level metrics are shown in blue. For disk usage and memory usage, the native host metric was transformed to have the same units as the container metric. All metrics have been transformed to the same units and scaled as a percentage of the maximum observed value. The four stages of the pipeline include downloading the data (download), splitting and demultiplexing the reads (split), aligning the reads to the reference (align), and assembling the counts while removing duplicate reads (merge). We observe that the container and VM-level metrics mostly overlap in the stages. However, there are differences when there are background processes, most notably when there is considerable disk usage. The alignment stage is also notable in that we can see that the CPU usage declines near the end, probably indicating that the pipeline is waiting on some slower threads (i.e. stragglers) to finish before it can proceed, indicating this stage might be improved with better load balancing, or with smaller workloads for the threads. This is an example of how the {\em Container Profiler} can be used to identify portions of the pipeline that can be optimized.} 
\centering
\includegraphics[width=\textwidth]{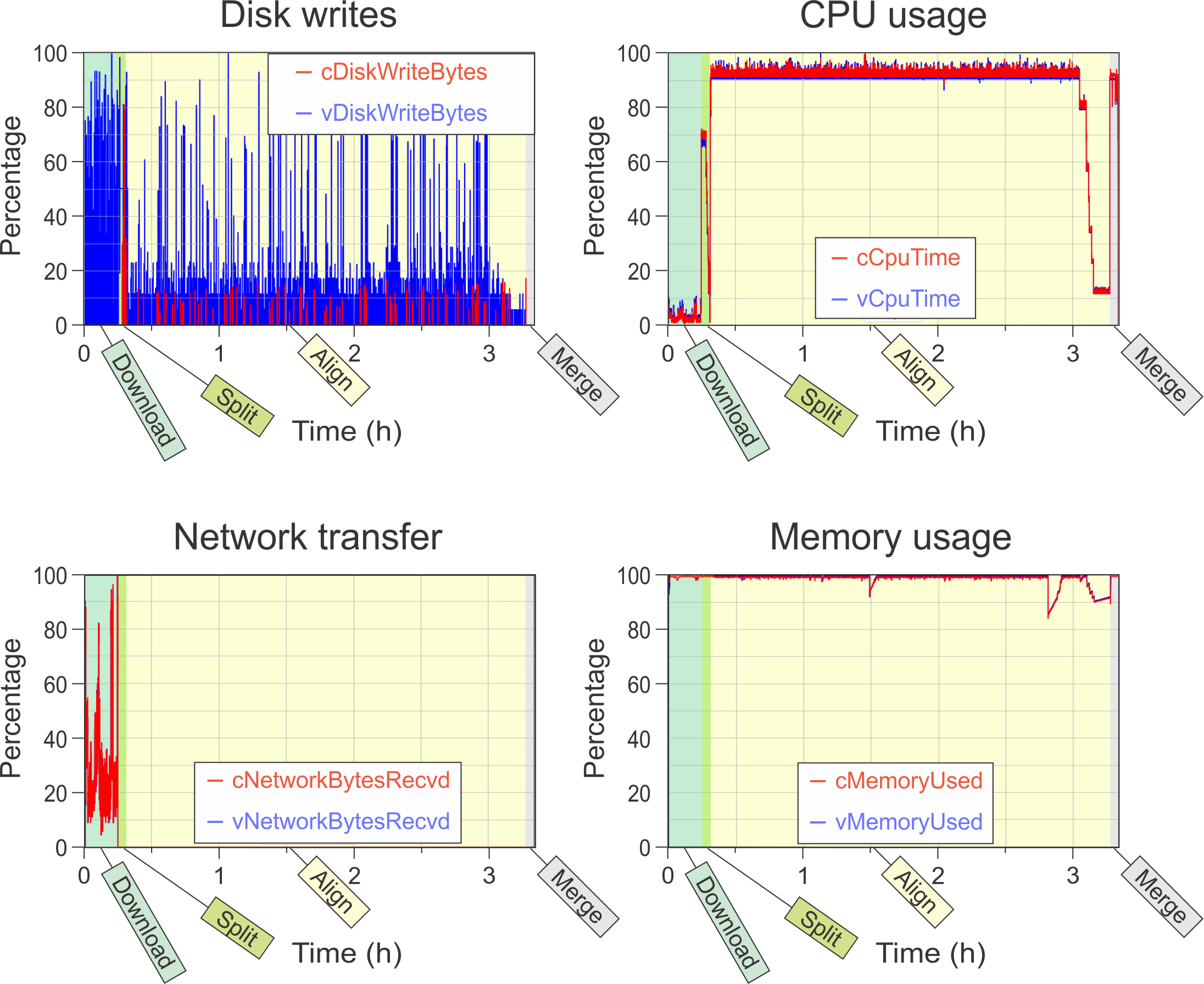}
\label{fig:benchmark}
\centering
\end{figure}

\subsection{{\em Container Profiler} can inform pipeline optimization}
Figure~\ref{fig:rugraphg} summarizes the CPU utilization characteristics of different stages of the UMI RNA-seq pipeline. The CPU usage profile is consistent with our expectations. The execution of the align and merge steps are expected to be bound by CPU resources and they indeed spent the majority of the time executing source code. Download is limited by the network bandwidth and the split stage by disk I/O. Hence the cpuIdle time is highest in these stages.

Despite the fact that the align stage is expected to be limited by the CPU resources, there is significant CPU-idle time during that stage. This suggests the presence of a bottleneck that may be the target for further optimization. We collected CPU, memory, network, and disk utilization metrics at both the container and VM/host levels for the RNA sequencing analytical pipeline. These are visualized in Figure~\ref{fig:benchmark}.  Note that the x-axis depicting time in this figure encompasses the entire pipeline incorporating all stages: download, split, align, and merge. Overall our profiling results depict resource utilization patterns that we expected. The download stage consumes network resources. The split stage is the most disk intensive step. The alignment and merge stages consume the most CPU resources. Our profiling data also points to areas where resource consumption may be a problem. For example, memory usage is high for all the stages. This may be due to greedy allocation by the executables, or it may indicate that allocating more memory could benefit the pipeline. Most interesting, is CPU utilization during the alignment stage. Just before the 3 hour mark, we see a series of drops over the next 30 minutes, creating a ladder of 8 steps. The alignment stage uses up to 8 vCPUs to align different files of reads simultaneously. Near the end of the alignment stage, most of the files will have been processed and there will be more available vCPUs than unprocessed files. As a result, the CPU utilization drops as vCPUs lie idle waiting for the final files to be processed. However, this under-utilization of resources lasts for 30 minutes indicating that the final files are rather large. This presents an opportunity to improve pipeline performance by splitting the processing into smaller files (which is an option in the split software), or by processing the largest files first. We would not have known about these potential optimizations without fine-grained profiling results from the {\em Container Profiler}.

\subsection{Container-level metrics can provide useful additional information}
A key feature of the {\em Container Profiler} is the ability to capture container-level metrics to describe resource utilization of only the containerized task(s). We expect these metrics to be similar, and that they could differ given that the VM/host level metrics also encompass resources being used by processes running on the host external to the container and pipeline. Since we only ran our pipeline on an dedicated test VM, the container metrics should be very similar to the VM/host metrics, which was in fact the case from our observations. However, one can see differences between the disk utilization metrics during the split and alignment stages where there are a large number of disk writes to the host file system. Docker manages these disk writes by providing the container with an internal mount point which is eventually written to a host file. The caching and management of this data is external to the container and is not captured by the container metrics, but is captured by the host metric. In addition, during the alignment stage, intermediate results from the aligner are continuously piped to another process which then re-formats the intermediate output and writes the final output to a file on the host system. Multiple threads are used, more than the available number of cores resulting in frequent context switches.  The pipe management and context-switching are also handled by the operating system and are captured by the host metric and not the container metrics. The separation of container and OS based consumption can be useful for example, when trying to assess effects due to resource contention that may occur when multiple jobs are run on the same physical host, which often happens on public clouds where the assignment of instances to hosts is controlled by the vendor.

\subsection{{\em Container Profiler} can sample container and host metrics with sub-second resolution}
\begin{figure}[H]
\caption{Distribution plot (log-scale) of time required to collect profiling data. We profiled resource utilization of the RNA-sequencing pipeline on an IBM Cloud bx2d-metal-96x384 virtual machine with dual Intel Platinum 8260 CPUs at 2.4 GHz, with 96 virtual CPU cores, 384GB of memory, and a 960 GB SATA M.2 mirrored SSD as the local boot disk). We executed the complete RNA-seq pipeline four times to profile 1) only VM/host metrics, 2) VM/host and container metrics, 3) \textit{ALL} metrics, and no metrics by running the pipeline in the absence of the profiler. Plots depict time to collect resource utilization samples at one-second intervals with the Container Profiler while running the entire RNA-seq pipeline. Time to \underline{collect ~9000 samples} of each type (Process-level, Container-level, and VM-level) is shown. 99.95\% of process-level samples were collected under 100 milliseconds, while all container-level samples were collected under 74 milliseconds, and all VM-level samples were collected at or under 60 milliseconds. The figure shows the process-level, container-level, and VM-level profiling time distribution over 120 milliseconds on the x-axis. The 90th percentiles for sample collection are shown.} 
\label{fig:percentile}
\centering
\includegraphics[scale=1.55]{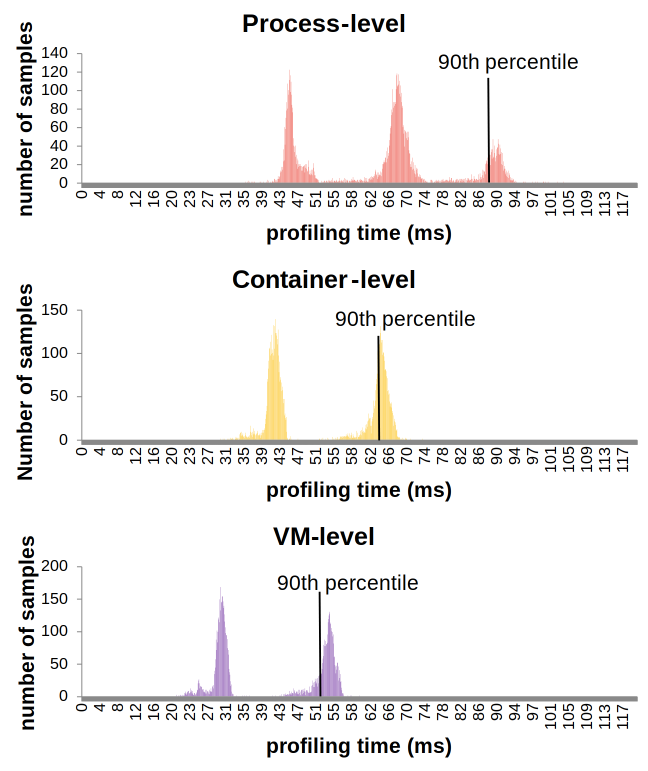}
\end{figure}

For the {\em Container Profiler} to be useful, the collection of profiling metrics must have sufficiently low overhead to enable rapid sampling of resource utilization to collect many samples for time series analysis. The time required to collect the metrics limits the granularity of the profile. To achieve 1 second sampling for time series analysis requires the ability to repeatedly sample resource utilization every 1 second (1000 ms). However, profiling time is not constant, and depends on the state of resources being utilized by the containerized pipeline and the host. The variability of profiling time is shown in the histogram in Figure ~\ref{fig:percentile}. When profiling our RNA-sequencing pipeline, VM-level and container-level profiling had a bi-modal distribution, while process-level sampling had a tri-modal distribution. The slowest profiling was observed during the stressful compute-bound alignment stage of the pipeline.

For all levels of profiling verbosity, the {\em Container Profiler} was able to profile resource utilization in less than 100ms. The longest profiling time and highest variation was for process-level profiling as metrics are collected for each process in the pipeline. The number of processes can vary throughout the execution of complex parallel pipelines, as was the case for the align stage of our RNA-sequencing pipeline. Our RNA-sequencing pipeline featured a maximum of 85 concurrent processes during the align stage. These processes ran for approximately 39\% of the duration of the align stage. The time required to capture host and container level metrics was less variable as the number of metrics collected is fixed. As shown in Figure ~\ref{fig:percentile}, 90\% of the time, the container and host level metrics were collected in less than ~63 milliseconds and always under 75 milliseconds. The process metrics do take longer to collect but still less than 100 milliseconds in the worst case.  Profiling at the process-level involves collecting all metrics every second.  For profiling our UMI RNA-sequencing pipeline use case which required ~2.5 hours to execute with one-second sampling and full profiling verbosity (process-level metrics), ~9,000 JSON files were collected which required 296 MB of storage space.

\subsection{{\em Container Profiler} has lower overhead than the variation in pipeline execution time on public clouds}

A design objective for the {\em Container Profiler} is to not significantly impact the performance of the pipeline being profiled. Failing to realize this objective may result in the overhead from resource profiling impacting the collected metrics. While some overhead is unavoidable, ideally it should be lower than the inherent variations of pipeline execution time on the public cloud. 


\begin{figure}[H]
\caption{This figure depicts the profiling overhead of the {\em Container Profiler} and the resulting percentage increase in the total runtime of the entire RNA-seq pipeline. The increases in run time are very modest: Host/VM only (0.07\%), Host/VM + Container (0.09\%), and Host/VM + Container + Process (0.71\%). Error bars depict one standard deviation from the average. Standard deviation of pipeline runtime for 5 runs of the RNA-seq pipeline on the IBM bx2d-16x64 Virtual Machine with no profiling was (1.38\%), approximately 194\% greater than the worst case overhead of the Container Profiler when profiling with full verbosity (i.e. collecting all metrics). }
\centering
\includegraphics[width=\textwidth]{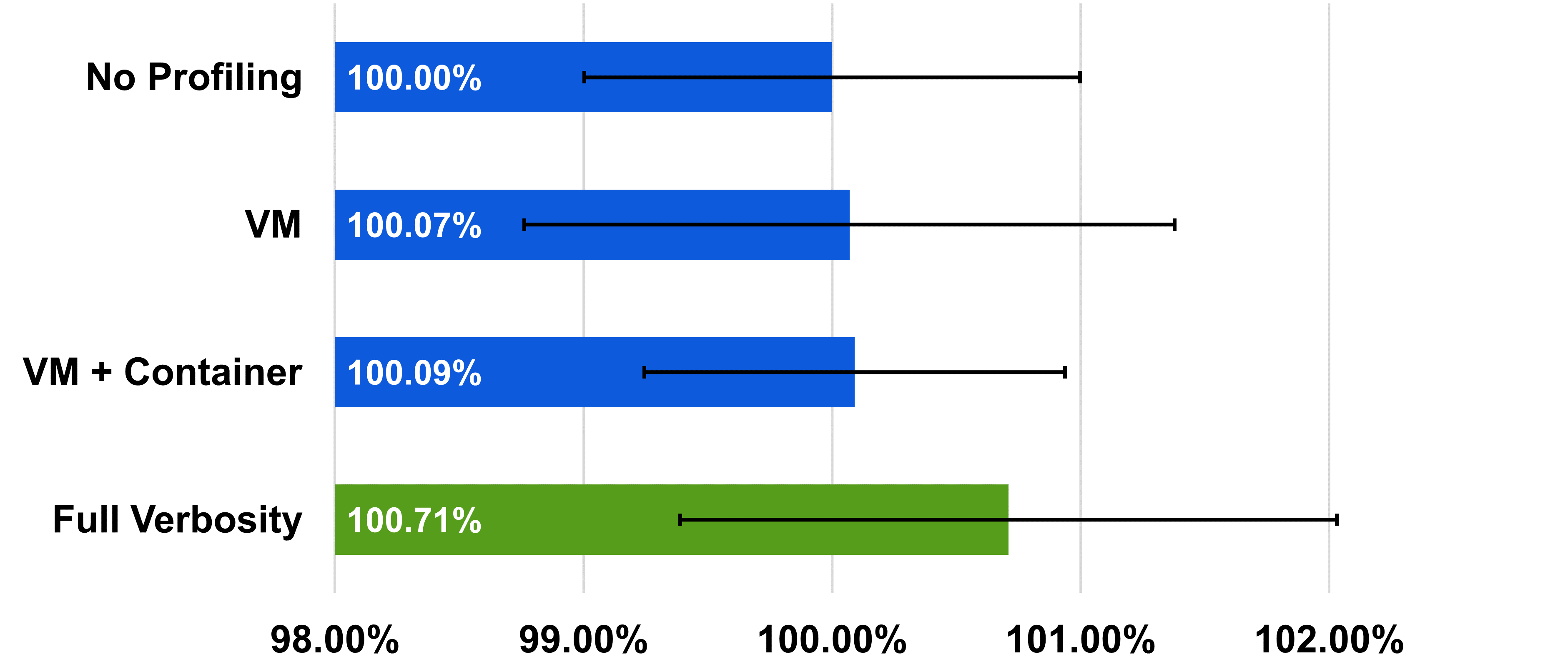}
\label{fig:ProfOH}
\centering
\end{figure}

To measure the performance impact of resource utilization profiling when running the RNA-seq pipeline, we initially attempted to assess the overhead using Amazon Elastic Compute Cloud (EC2) cloud VMs.  However, we discovered that the runtime of the RNA-seq pipeline varied by more than 5\% on Amazon EC2, which was more than 5x greater than the overhead of the {\em Container Profiler}. This degree of performance variance made it difficult to evaluate the performance overhead of the {\em Container Profiler} since we could not easily distinguish between pipeline performance variance and profiling overhead on EC2. We then measured the performance overhead of the {\em Container Profiler} by profiling the pipeline using the IBM cloud bx2d-metal-96x384 server which had performance variance around ~1\%. Figure~\ref{fig:ProfOH} depicts the overhead from one-second resource utilization sampling by the {\em Container Profiler} for the RNA-seq pipeline on the IBM metal server.  IBM metal servers are private and not shared with multiple users. Running on this isolated server greatly reduced the performance variance of running RNA-seq. We measured worst case overhead for the {\em Container Profiler} to be 0.71\%, which equates to about 3.4 minutes for an 8-hour pipeline with full verbosity metrics collection (VM + container + process).  Overhead is reduced to as little as .07\% overhead, or about 20 seconds for an 8-hour pipeline when only collecting VM-level metrics.  Adding container-level, and especially process-level metrics slightly increased the runtime of the RNA-seq pipeline for collecting resource utilization data.  We believe that this profiling overhead is within an acceptable level and note that even at maximum profiling verbosity, it is substantially less than the observed performance variance for running our RNA-seq pipeline on a public cloud VM. Users can reflect on our reported overhead times to make informed decisions when planning to profile their own pipelines.

\section{Methods}

\subsection{Implementation Details}

\begin{figure}[H]
\caption{Summary of Bash scripts used in the implementation of Container Profiler.}
\centering
\includegraphics[width=\textwidth]{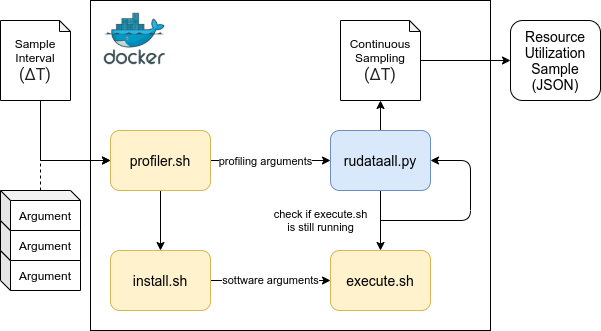}
\label{fig:CPWorflow}
\centering
\end{figure}

The {\em Container Profiler} is implemented as a collection of Bash and Python scripts. Figure~\ref{fig:CPWorflow} provides an overview. There are three basic use cases for building a docker image for the Container Profiler. The first use case allows users to profile an existing Docker container by providing a Dockerfile which specifies their own setup and software installation inside the container. This is the simplest approach to profiling when the user has a working Dockerfile. The other two use cases support users who do not know how to write Docker files but are familiar with writing Bash scripts. The second use case gives users the ability to install all software inside the Docker image when the software installation becomes too complicated to put in the Dockerfile. In other words, it puts the required installation commands into a script that will be executed by the Dockerfile. For the third use case, the user provides their own executable bash script as the entry point in the Docker container. This use case can help the user simplify a set of commands they have to profile. In this case, the user just puts a set of commands into an executable script file and runs it as the entrypoint of the container. When the {\em Container Profiler} is executed inside a Docker container, it snapshots the resource utilization for the host (i.e. VM), container, and all processes running inside the container producing output statistics to a \texttt{.json} file. A sampling interval (e.g. once per second) is specified to configure how often resource utilization data is collected to support time series analysis of containerized applications and pipelines. Time series data can be used to train mathematical models to predict the runtime or resource requirements of applications and pipelines. Time series data can be visualized by using matplotlib Python graphing scripts that are included with the {\em Container Profiler}.

To improve the periodicity of time series sampling, we continuously subtract the most recent observed run time of the {\em Container Profiler} for sample collection from the configured sampling interval (e.g. 1 second) in {\tt rudataall.py}.  This approach notably improved the periodicity of sampling when the container was under load improving our ability to obtain samples at evenly spaced intervals. To enable addressing any potential drift of sample collection times, we capture timestamps for when each resource utilization metric is sampled in the output JSON.  These timer ticks enable precise calculation of the time that transpires between resource utilization samples for each metric. This allows the rate of consumption of system resources (e.g. CPU, memory, disk/network I/O) to be precisely determined throughout the pipeline's execution.
%
The {\em Container Profiler} consists of the profiling script and two supporting scripts (for installation and pipeline execution) depicted in Figure~\ref{fig:CPWorflow}: {\tt profiler.sh}, {\tt install.sh}, and {\tt execute.sh}.

The {\tt profiler.sh} script is the primary script that generates profiling information in JSON format describing resource utilization of the containerized task. The {\tt profiler.sh} script requires the user to provide a command or a set of commands along with arguments to start the profiling. This script internally invokes another Python script {\tt rudataall.py}.

The {\tt rudataall.py} script collects the resource utilization data. Specifically, this script takes a snapshot of the resource utilization metrics and records output to a JSON file using the time of the sample as a unique filename.  The script accepts parameters {\tt -v}, {\tt -c}, and {\tt -p} to inform the tool what type of data to collect: VM, container, and/or process\-level metrics respectively.  The default behavior when running this script without any parameters is to collect all metrics.

The {\tt profiler.sh} script only works if the workflow/software is already installed in the containerized environment. This means that we cannot profile workflows/software that has not been containerized. The {\em Container Profiler} provides an option that enables users to install software in a container using the {\tt install.sh} script. Users provide a set of commands in the {\tt install.sh} script to install their dependencies and software they wish to profile. Once installed, the user can run the {\tt profiler.sh} script against the newly installed software. To profile resource utilization of a bash script, users can specify a series of commands using the optional {\tt execute.sh} script to configure profiling.

Some users may be more familiar with editing Dockerfiles instead of bash scripts. We provide support for users to provide their own Dockerfile to build a custom container to be profiled.

\subsection{Technical details using our scripts}
To use the {\em Container Profiler} scripts with any container, a Linux based Docker container that encapsulates a script or job to run inside is required. To configure the {\em Container Profiler} tool to profile the container, users can optionally provide an executable script inside the {\em Container Profiler} which is specified during the {\tt build.sh} script. In the executable script, the user launches the container's job or task to be profiled.

The {\tt profiler.sh} script has four different modes: {\tt profile}, {\tt delta}, {\tt csv}, and {\tt graph}:

\begin{table*}[h!t]
  \renewcommand{\arraystretch}{1.3}
  \caption{Container Profiler with four different modes}
  \label{tab:modes}
  \begin{tabular}{l|l}
    \hline 
    \textbf{Mode} & \textbf{Description} \\
    \hline
    \hline
    profile & profile resource utilization \\
    \hline
    delta & calculate the different between two profiling samples \\
    \hline
    csv & convert a set of JSON resource utilization files into a single CSV file \\
    \hline
    graph & generate profiling graph(s) from a CSV file \\
    \hline
  \end{tabular}
\end{table*}

For the {\tt profile} mode, there are two required parameters: the output directory specifies the location of generated profiling files in JSON format, and the time interval specifies a time series sampling interval in milliseconds. The profiler generates a JSON file at the beginning and the end of the process if the sampling interval is set to zero. Otherwise, the profiler generates a JSON file at each sampling interval. The Container Profile also collects static metrics which typically describe hardware characteristics. The profiler first checks if a static information file exists (static.json). If missing, the profiler captures static parameters and writes out the static information file at the start of profiling. By default 11 static metrics are captured. They include: the host's kernel info, the host's cpu type, CPU Level 1 instruction cache size, CPU Level 1 data cache size, CPU Level 2 cache size, CPU Level 3 cache size, host boot time, host VM ID, the number of CPU cores available to the container, and the container ID.

For the {\tt delta} mode, there are two required parameters: the input directory which contains the original raw JSON files, and the output directory where the delta JSON files will be written. The {\tt delta| mode} also provides an option to allow users to specify the modification operator for performing the delta. The default delta operator calculates the difference between two samples (i.e. final minus initial value). The typical use case is to calculate the delta of the resource utilization between the first and last sample to capture the full resource utilization of a task or pipeline. Other operators include max, min, and average to determine the max, min, and average values of metrics from a set of JSON files. 

For the {\tt csv} mode, there are two required parameters: the input directory that contains processed JSON files in delta format, and the name of an output CSV file where all resource utilization data from the processed JSON files will be aggregated to.

For the {\tt graph} mode, there are two required parameters: the input CSV file capturing all resource utilization data from processed JSON files, and the output directory for writing graph files. In addition, there are a few other options such as one to specify whether to plot the curves together or using separate graph files.

\subsection{Visualization}
The {\em Container Profiler} in the {\tt graph} mode also provides an option to specify the creation of time-series graphs. The graphing configuration file supports multiple settings to specify how to generate graph(s). Each graph configuration file should start with a line that includes the components: the {\tt \#\#\#} followed by the title and the y-coordinate label. This is followed by line(s) that describe the metric(s) that users want to output in a single graph (one metric per line). As a starting point, a default graph configuration file {\tt graph.cfg} is provided in the {\tt cfg} directory.

\section*{Availability of supporting data and materials}

\begin{itemize}
\item Project name: Container Profiler
\item Project home page: \url{https://github.com/wlloyduw/ContainerProfiler} 
\item Contents available for download: Docker Images, Dockerfiles, installation scripts, and execution scripts. 
\item Operating system(s): Linux, Mac OS X, Microsoft Windows. 
\item Programming language(s): Bash, Python
\item License: MIT License
\end{itemize}

\section*{List of abbreviations}
\noindent
AWS: Amazon Web Services; EC2: Elastic Compute Cloud; VM: virtual machine; CPU: central processing unit; IaaS: Infrastructure-as-a-Service; RNAseq: RNA sequencing; LINCS: Library of Integrated Network-Based Cellular Signatures; DToxS: Drug Toxicity Signature; RNA: ribonucleic acid; cgroup: container control group.

\section*{Competing Interests}
LHH and KYY have equity interest in Biodepot LLC, which receives compensation from NCI SBIR contract numbers 75N91020C00009 and 75N91021C00022. The terms of this arrangement have been reviewed and approved by the University of Washington in accordance with its policies governing outside work and financial conflicts of interest in research. 

\section*{Author's Contributions}

VH, LHH, HD, RS, and DP contributed to the development of the Container Profiler.  LHH implemented Docker containers for RNA-seq pipelines. VH, RS, NA, and DP conducted performance testing and empirical experiments. KYY, RS, WL, VH, and LHH drafted
the manuscript. WL, KYY, and LHH designed the case study. WL
provided cloud computing expertise. WL and KYY coordinated the benchmarking experiments. All authors edited the manuscript.

\section*{Acknowledgements}
LHH, HD, RS, WL, and KYY are supported by the National Institutes of Health (NIH) grant R01GM126019.  DP is supported by the NIH Diversity Supplement R01GM126019-02S2. LHH and KYY are also supported by NIH grants U24HG012674 and R03AI159286. WL is also supported by NSF grant OAC-1849970.  We acknowledge support from the AWS Cloud Credits for Research and IBM Cloud Credits (awarded to LHH, WL, and KYY).

%
%

\printbibliography

\end{document}